\documentstyle[pre,aps]{revtex}
\begin{document}
\twocolumn[\hsize\textwidth\columnwidth\hsize
           \csname @twocolumnfalse\endcsname
\title{Non-instant collisions and two concepts of quasiparticles}
\author{Pavel Lipavsk\'y and V\'aclav \v Spi\v cka}
\address{Institute of Physics, Academy of Sciences, Cukrovarnick\'a 10,
16200 Praha 6, Czech Republic}
\author{Klaus Morawetz}
\address{Fachbereich Physik, University Rostock, D-18055 Rostock,
Germany}
\pacs{
\Pacs{05}{30$-$d}{Quantum statistical mechanics}
\Pacs{67}{55$-$s}{Normal phase of liquid $^3$He}
\Pacs{03}{65Nk}{Nonrelativistic scattering theory}
\Pacs{25}{70$-$z}{Low and intermediate energy heavy-ion reactions}
}
\maketitle
\begin{abstract}
The kinetic theory recently implemented in heavy ion reactions combines
a non-local and non-instant picture of binary collisions with
quasiparticle features. We show that the non-instant description is
compatible with the spectral concept of quasiparticles while the
commonly used variational concept is consistent only with instant
collisions. The rearrangement energy, by which the variational concept
surpasses the spectral one, is shown to be covered by a medium effect
on non-instant collisions.
\end{abstract}
\vskip2pc]

The quasiparticle concept provides a basic theoretical framework for
description of interacting Fermi systems. Although quasiparticles are
well defined only close to the ground state, i.e. at small temperatures
and under weak perturbing fields, a lack of tractable theories for
systems far from equilibrium forces physicists to deal with
quasiparticles also in this region. From a number of highly
non-equilibrium systems treated in the quasiparticle picture one should,
perhaps, mention heavy ion reactions where the excitation energy far
exceeds the Fermi energy.

It is disturbing that an extension of the quasiparticle concept far
from the ground state is not unique. There are two formulations of the
quasiparticle concept, the phenomenologic and the microscopic. The
former relies on the variation of thermodynamic quantities, the latter
on properties of the single-particle spectrum. Close to the ground state
these concepts are proven to be identical but they become increasingly
distinct as an excitation of the system increases. A natural question
arises: Which concept of quasiparticles works better in kinetic
equations for highly non-equilibrium systems? As we show in this paper,
the answer depends on approximations employed in the collision integral.
If the collisions are treated as local and instant, as it is common in
equations of Boltzmann type, the variational concept seems to be better,
at least with respect to processes of a thermodynamic and hydrodynamic
character. In contrary, a non-instant treatment of collisions, as
suggested by Danielewicz and Pratt \cite{DP96}, is compatible only with
the spectral concept.

Let us remind first both concepts. In the original phenomenologic
formulation, Landau \cite{L56,AGD63} postulates that a density of
quasiparticles equals the density of composing particles,
\begin{equation}
n=\int{dk\over(2\pi)^3}\tilde f_k,
\label{e1}
\end{equation}
where $\tilde f_k$ is the quasiparticle distribution in momentum space.
Assuming furthermore that quasiparticles cover all degrees of freedom of
the system, the quasiparticle energy $\tilde\epsilon$ can be defined as
a variation of the total energy $\cal E$,
\begin{equation}
\tilde\epsilon_k={\delta{\cal E}\over\delta\tilde f_k}.
\label{e2}
\end{equation}
From the entropy follows that the equilibrium distribution is of the
Fermi-Dirac form, $\tilde f_k=f_{FD}(\tilde\epsilon_k-\tilde\mu)$. Out
of equilibrium, $\tilde f_k$ differs from the Fermi-Dirac distribution
and relations (\ref{e1}-\ref{e2}) hold locally in time and space.

The microscopic Green's function foundations \cite{AGD63,NL62,LN62} of
the quasiparticle theory define the quasiparticle energy alternatively
from a singularity of the single-particle spectral function
\begin{equation}
A={\Gamma\over\left(\omega-{k^2\over 2m}-\Sigma\right)^2+
{1\over 4}\Gamma^2},
\label{e3}
\end{equation}
where $\Sigma$ and ${1\over 2}\Gamma$ are real and imaginary parts of
the selfenergy. For long-living excitations, $\Gamma\to 0$, the spectral
definition of the quasiparticle energy reads
\begin{equation}
\epsilon_k={k^2\over 2m}+\Sigma(\epsilon_k,k).
\label{e4}
\end{equation}
The quasiparticle distribution is identified from a singularity of the
single-particle correlation function $G^<$. In equilibrium, where
$G^<(\omega,k)=f_{FD}(\omega)A(\omega,k)$, this approach also yields the
Fermi-Dirac distribution $f_k=f_{FD}(\epsilon_k-\mu)$. Again, out of
equilibrium all relations are locally valid, except for the Fermi-Dirac
form of $f_k$.

Various studies \cite{AGD63,L59,C66} prove that close to the ground
state Landau's quasiparticles are the elementary excitations seen in
the single-particle spectrum. In these proofs it is essential that
dissipative processes freeze out with the square of the temperature so
that quasiparticles become free of collisions. Far from the ground
state, however, the collisions become important. As long as the
collisions are local and instant, as it is the case for a weak
interaction, they have no effect on the thermodynamic properties and
both quasiparticle concepts remain equivalent. When non-local,
the collisions affect thermodynamic properties in a way which escapes
the variational approach. A clear-cut example is the system of hard
spheres whose density and total energy do not differ from the ideal
gas and yet its equation of state includes the virial corrections know
as the unaccessible volume. The non-instant collisions result in even
deeper changes on which we focus in this paper.

We will discuss two groups of virial corrections due to which the
spectral quasiparticles do not satisfy postulates of the
variational concept. Both are caused by the finite duration of
collisions. First, the density of spectral quasiparticles differs
from the density of composing particles by the correlated density,
$n^{\rm cor}=n-\int{dk\over(2\pi)^3}f_k$, as it is known from the
Beth-Uhlenbeck virial expansion \cite{SRS90}. Second, the
variational quasiparticle energy $\tilde\epsilon$ differs from
the spectral one by the rearrangement energy, $\epsilon^{\rm re}=
\tilde\epsilon-\epsilon$. This has been observed earlier \cite{GH83}
for the Galitskii--Feynman approximation widespread used either to
evaluate the total energy ${\cal E}$ as the starting point of the
variational approach or to evaluate the selfenergy needed in the
spectral approach. It should be noted that the Galitskii-Feynman
approximation is missing the particle-hole channels important at
very low temperatures. Our discussion is thus limited to rather
highly excited systems.

Collisions of quasiparticles has been recently studied in \cite{SLM98}
with the help of methods developed for gases \cite{B69,BKKS96}. A
kinetic equation derived from non-equilibrium Green's functions by a
systematic gradient expansion includes a scattering integral in which
collisions are described as non-local and non-instant events. Although
small in slowly varying systems, these non-local and non-instant
corrections appreciably influence a behavior of the system since they
affect conserving quantities and therefore contribute to thermodynamic
and hydrodynamic properties. For instance, the non-locality corrects for
the unaccessible volume and the finite duration yields the correlated
density. On top of these two effects known from the theory of gases,
the found collisions possess a new feature in that the momentum and the
energy of a colliding pair of quasiparticles do not conserve. Small
amounts of momentum and energy are gained by a colliding pair due to
changes of the Pauli blocking during the collision.

The transfer of momentum and energy between the colliding pair and the
medium of background particles provides an important link between the
variational and the spectral concepts of quasiparticle energies. Since
we want to focus on this energy balance, we assume a homogeneous system
for simplicity. We will show that the rearrangement energy contributing
to the variational energy simulates for the energy gained by a pair of
quasiparticles during in-medium collisions.

The variational approach works only if binary collisions are treated
within the instant and elastic approximation. To show why, assume a
phenomenologic kinetic equation
\begin{equation}
{\partial\tilde f_k\over\partial t}=\tilde I_k,
\label{e5}
\end{equation}
where $\tilde I_k$ is an unspecified collision integral. To conserve the
number of particles,
\begin{equation}
{dn\over dt}=\int{dk\over(2\pi)^3}{\partial\tilde f_k\over\partial t}=
\int{dk\over(2\pi)^3} \tilde I_k,
\label{e6}
\end{equation}
the collision integral has to satisfy $\int dk\tilde I_k=0$ which is
possible only for instant collisions. In addition, from the energy
balance,
\begin{equation}
{d{\cal E}\over dt}=\int{dk\over(2\pi)^3}
{\delta{\cal E}\over\delta\tilde f_k}{\partial\tilde f_k\over\partial t}
=\int{dk\over(2\pi)^3}\tilde\epsilon_k\tilde I_k,
\label{e7}
\end{equation}
it follows that the total energy conserves, ${d{\cal E}\over dt}=0$,
only if the sum of variational quasiparticle energies conserve in
collisions, $\int dk\tilde\epsilon_k\tilde I_k=0$.

In contrast, the kinetic equation resulting from the spectral concept
as an asymptotic of non-equilibrium Green's function in the
Galitskii--Feynman approximation \cite{SLM98},
\begin{eqnarray}
{\partial f_k\over\partial t}
&=&\int{dpdq\over(2\pi)^5}\delta(\epsilon_k+\epsilon_p-\epsilon_{k-q}^--
\epsilon_{p+q}^--2\Delta_E)
\nonumber\\
&&\times |T^-|^2(1-f_k-f_p)f_{k-q}^-f_{p+q}^-
\nonumber\\
&-&\int{dpdq\over(2\pi)^5}\delta(\epsilon_k+\epsilon_p-\epsilon_{k-q}^+-
\epsilon_{p+q}^++2\Delta_E)
\nonumber\\
&&\times |T^+|^2f_kf_p(1-f_{k-q}^+-f_{p+q}^+),
\label{e8}
\end{eqnarray}
has a non-instant collision integral which does not conserve the
sum of quasiparticle energies. For distributions we use abbreviations
$f=f(t)$ and $f^\pm=f(t\pm\Delta_t)$, and similarly for arguments of
quasiparticle energies. The T-matrix, $T_R=|T|{\rm e}^{i\phi}$, is
centered between asymptotic states, $T^\pm=T^\pm(\epsilon_1^\pm+
\epsilon_2^\pm\pm\Delta_E,k,p,q,t\pm{1\over 2}\Delta_t)$. Apparently,
the collision delay, $\Delta_t=\left.{\partial\phi\over\partial\Omega}
\right|_{\Omega=\epsilon_1+\epsilon_2}$, and the energy gain of a
colliding pair, $2\Delta_E=-\left.{\partial\phi\over\partial t}\right|_
{\Omega=\epsilon_1+\epsilon_2}$, do not meet phenomenologic expectations
about the structure of the collision integral.

The difference between (\ref{e5}) and (\ref{e8}) is even more obvious
from the conservation laws found in \cite{SLM98} from (\ref{e8}) by
integration over momentum $k$ with factors 1 and $\epsilon_k$. The
balance of the number of particles,
\begin{equation}
{dn\over dt}={d\over dt}\int{dk\over(2\pi)^3}f_k+
{d\over dt}\int dP\Delta_t ,
\label{e9}
\end{equation}
where
\begin{eqnarray}
dP&=&{dkdpdq\over(2\pi)^8}
|T|^2\delta(\epsilon_k+\epsilon_p-\epsilon_{k-q}-\epsilon_{p+q})
\nonumber\\
&&\times f_kf_p(1-f_{k-q}-f_{p+q}),
\label{e10}
\end{eqnarray}
includes the term proportional to the collision delay. This is
exactly the correlated density, $n^{\rm cor}=\int dP\Delta_t$,
found in \cite{SRS90} from the equilibrium Green's functions. The
kinetic equation (\ref{e8}) thus implies that the number of spectral
quasiparticles does not equal the number of composing particles.

The energy balance found from (\ref{e8}),
\begin{equation}
{d{\cal E}\over dt}=\int\!{dk\over(2\pi)^3}\epsilon_k
{\partial f_k\over\partial t}+{d\over dt}\int\!dP
{\epsilon_k+\epsilon_p\over 2}\Delta_t-\int\!dP\Delta_E,
\label{e11}
\end{equation}
also includes contributions that are not compatible with the
phenomenologic postulates. Similarly to the correlated density,
there is the energy of correlated particles $\propto\Delta_t$.
Moreover, due to the transfer between the background and the
colliding pair, there is a mean energy gain $\propto\Delta_E$
by which the energy covered by single-particle degrees of freedom
(the sum of quasiparticle energies) can change in time.

One might wonder whether the kinetic equation (\ref{e8})
conserves the energy at all because the right hand side of balance
equation (\ref{e11}) does not have a transparent form of the total time
derivative. Although it is a tedious task, it can be shown that
(\ref{e8}) conserves the energy given by
\begin{eqnarray}
{\cal E}&=&\int\!{dk\over(2\pi)^3}f_k{k^2\over 2m}+{1\over 2}\!
\int\!{dkdp\over(2\pi)^6}f_kf_p
{\rm Re}T_R(\epsilon_k\!+\!\epsilon_p,k,p,0)
\nonumber\\
&+&
\int dP{\epsilon_k+\epsilon_p\over 2}\Delta_t.
\label{e12}
\end{eqnarray}
One can check that (\ref{e11}) results from the time derivative of
(\ref{e12}). The energy of correlated particles directly corresponds to
the second term of (\ref{e11}). The second term of (\ref{e12}) splits
into the selfenergy part of the quasiparticle energy and into the mean
energy gain. The mean energy gain follows exclusively from the time
derivative of ${\rm Re}T_R$ in agreement with fact that the effect of
medium on the binary collision is responsible the energy transfer
between colliding particles and the background. We note that the total
energy (\ref{e12}) is identical with the Galitskii--Feynman
approximation in the limit of small scattering rates.\footnote{
Although formula (\ref{e12}) holds out of equilibrium, we want to
outline its simple derivation for the equilibrium case. In the general
expression for the energy \cite{KB62},
$${\cal E}=\int{dkd\omega\over(2\pi)^4}{1\over 2}
\left(\omega+{k^2\over 2m}\right)f_{FD}(\omega)A(\omega,k),$$
one employs the limit of small scattering rates
\cite{C66,SRS90,SLM98,BKKS96,KM93},
$$A=\left(1+{\partial\Sigma\over\partial\omega}\right)
2\pi\delta(\omega-\epsilon_k)+
{\rm Re}{\Gamma\over(\omega-\epsilon_k+i0)^2}.$$
The second term represents off-shell contributions neglected within the
so-called quasiparticle approximation commonly used to derive the
quasiparticle theory from Green's functions. Its inclusion is essential
for all correlated quantities.}

Being able to cover the correlated density and the latent heat
due to the mean energy gain, the spectral concept provides a more
sophisticated description of interacting Fermi liquids than the
phenomenological one. This superiority, however, goes on the cost of
such complex corrections as the collision delay and energy gain
during collisions. It is interesting to see under what conditions the
theory based on the spectral concept reduces to the phenomenologic one.

According to (\ref{e6}), the phenomenologic concept works only if
the collision duration is negligible. Sending $\Delta_t\to 0$ in
(\ref{e8}) one obtains the instant collision integral and consequently
no correlated density appears in the number of particles balance,
$\int dP\Delta_t\to 0$. At least with respect to the density of
quasiparticles one can say that the distribution of quasiparticles
becomes close to the variational distribution,
\begin{equation}
f_k\to \tilde f_k.
\label{e16}
\end{equation}
In the energy conservation (\ref{e11}) and the total energy (\ref{e12}),
the neglect of the collision delay suppresses the contribution of the
correlated particles, $\int dP(\epsilon_k+\epsilon_p)\Delta_t\to 0$. The
total energy which conserves within the instant but still non-elastic
approximation of collisions, has the familiar form \cite{FW71},
\begin{equation}
{\cal E}\to\!\int\!{dk\over(2\pi)^3}\tilde f_k{k^2\over 2m}+{1\over 2}\!
\int\!{dkdp\over(2\pi)^6}\tilde f_k\tilde f_p
{\rm Re}T_R\!(\!\epsilon_k\!+\!\epsilon_p,k,p,0).
\label{e17}
\end{equation}
which is commonly used as a starting point in variational approaches
\cite{GH83}.

In spite of the above similarities, it would be premature to conclude
that in the limit of instant collisions, $\Delta_t\to 0$, the spectral
and the variational approaches are identical. The energy balance
(\ref{e11}) in the instant approximation,
\begin{equation}
{d{\cal E}\over dt}\to\int{dk\over(2\pi)^3}\epsilon_k
{\partial\tilde f_k\over\partial t}-\int dP\Delta_E,
\label{e18}
\end{equation}
still includes the mean energy gain $\int dP\Delta_E$. There is
simple but incorrect argument that the mean energy gain can be
neglected as $\Delta_t\to 0$ (The energy gain follows from the
time-dependency of scattering phase shift, i.e. at the end, from
the time-dependency of a distribution of particles in the
background. In the instant the background particles have to time
to pass any energy to the colliding pair.). A neglect $\Delta_E\to 0$
leads to an inconsistency between the energy conservation (\ref{e18})
and the total energy (\ref{e17}). Indeed, the total energy (\ref{e17})
still includes ${\rm Re}T_R$ from which the mean energy gain arises.
A consistent elastic approximation of collisions thus cannot be
achieved by a simple neglect of the non-elasticity.

The link between the variational and the spectral concepts can be
established if one is concerned with processes in which global
conservation laws play the dominant role while individual
trajectories of quasiparticles are of a marginal importance. In
this hydrodynamic regime, it is possible to rearrange the mean
energy gain into a mean-field-like contribution to the quasiparticle
energy, $\epsilon_k^\Delta$. From a demand
\begin{equation}
\int{dk\over(2\pi)^3}\epsilon_k^\Delta
{\partial\tilde f_k\over\partial t}=-\int dP\Delta_E,
\label{e19}
\end{equation}
and a variational form of the energy gain,
\begin{equation}
\Delta_E=-{1\over 2}{\partial\phi\over\partial t}=\int{dk\over(2\pi)^3}
{\delta\phi\over\delta\tilde f_k}{\partial\tilde f_k\over\partial t},
\label{e20}
\end{equation}
one finds the mean-field-like correction as
\begin{equation}
\epsilon_k^\Delta={1\over 2}\int dP{\delta\phi\over\delta\tilde f_k}.
\label{e21}
\end{equation}
After a substitution of (\ref{e19}) into (\ref{e18}), a comparison with
(\ref{e7}) shows that the mean-field-like correction is exactly the
rearrangement energy, $\epsilon_k^\Delta=\epsilon^{\rm re}_k$. From
(\ref{e21}) follows that the rearrangement energy describes the effect
of the time-dependent Pauli blocking on the scattering phase shift
$\phi$.

Relation (\ref{e19}) shows an advantage of the variational concept.
The elastic collision integral, $\propto\delta
(\tilde\epsilon_k+\tilde\epsilon_p-\tilde\epsilon_{k-q}-\tilde\epsilon_
{p+q})$ with the variational quasiparticle energy $\tilde\epsilon=
\epsilon+\epsilon^{\rm re}$, yields the same energy conservation as the
non-elastic one, $\propto\delta(\epsilon_k+\epsilon_p-\epsilon_{k-q}-
\epsilon_{p+q}-2\Delta_E)$ with the spectral quasiparticle energy.
Without a sacrifice of the energy conservation, one can thus circumvent
an inconvenience of non-elastic collision integrals by an incorporation
of the rearrangement energy.

In summary, we would like to stress the answer on the question which of
the quasiparticle concepts is more suitable for highly non-equilibrium
systems. The spectral concept offers the more elaborate description of
the system dynamics provided that the collision integral includes the
collision delay and the energy gain. The instant and elastic picture of
collisions cannot capture such features of binary correlations like the
correlated density. The mean energy gain, however, can be mimicked by
the rearrangement energy included in the variational concept.

\medskip

Authors are grateful to Th.~Bornath, P.~Hub\'\i k, D.~Kremp and
J.~Kri\v stofik for useful discussions. This work was supported from the
Grant Agencies of the Czech Republic under contracts Nos.~202960098 and
202960021, and of the Czech Academy of Sciences under contract
Nr.~A1010806, the BMBF (Germany) under contract Nr.~06R0884 and the
Max-Planck-Society.


\begin{references}
\bibitem{DP96}
P. Danielewicz and S. Pratt,
Phys. Rev. C {\bf 53}, 249 (1996).
\bibitem{L56}
L. Landau,
Soviet Phys. -- JETP {\bf 3}, 920 (1956);
{\bf 5}, 101 (1957).
\bibitem{AGD63}
A. A. Abrikosov, L. P. Gorkov and I. E. Dzyialoshinski,
{\it Methods of Quantum Field Theory in Statistical Physics},
(Prentice Hall, New York, 1963).
\bibitem{NL62}
P. Nozi\`eres and J. M. Luttinger,
Phys. Rev. {\bf 127}, 1423 (1962)
\bibitem{LN62}
J. M. Luttinger and P. Nozi\`eres,
Phys. Rev. {\bf 127}, 1431 (1962).
\bibitem{L59}
L. D. Landau,
Soviet Phys. -- JETP {\bf 35}, 70 (1959).
\bibitem{C66}
R. A. Craig,
Ann. Phys. (NY) {\bf 40}, 416 (1966);
{\bf 40}, 434 (1966).
\bibitem{HCB64}
J. O. Hirschfelder, F. C. Curtiss and R. B. Bird,
{\it Molecular Theory of Gases and Liquids},
(Wiley, New York, 1964).
\bibitem{SRS90}
M. Schmidt, G. R\"opke and H. Schulz,
Ann. Phys. (NY) {\bf 202}, 57 (1990).
\bibitem{GH83}
H. R. Glyde and S. I. Hernadi,
Phys. Rev. B {\bf 28}, 141 (1983).
\bibitem{SLM98}
V. \v Spi\v cka, P. Lipavsk\'y and K. Morawetz,
Phys. Lett. A {\bf 240}, 160 (1998).
\bibitem{B69}
K. Baerwinkel,
Z. Naturforsch. a {\bf 24}, 22 (1969);
{\bf 24}, 38 (1969).
\bibitem{BKKS96}
Th. Bornath, , D. Kremp, W. D. Kraeft and M. Schlanges,
Phys. Rev. E {\bf 54}, 3274 (1996).
\bibitem{KB62}
L. P. Kadanoff and G. Baym,
{\it Quantum Statistical Mechanics},
(Benjamin, New York, 1962).
\bibitem{KM93}
H. S. K\"ohler and R. Malfliet,
Phys. Rev C {\bf 48}, 1034 (1993).
\bibitem{FW71}
A. L. Fetter and J. D. Walecka,
{\it Quantum Theory of Many-Particle Systems},
(McGraw-Hill, New York 1971), Chap. 4.
\end{references}
\end{document}